\documentclass[aps,preprint,groupedaddress]{revtex4}

\usepackage{graphics,epsfig}

\begin{document}

\title{The upper bound of the lightest Higgs boson in a Supersymmetric $SU(3)_L \otimes U(1)_X$ gauge model}

\author{R. Martinez}
\email[romart@ciencias.unal.edu.co]{}
\author{N. Poveda}
\email[npoveda@ciencias.unal.edu.co]{}
\author{J-Alexis Rodriguez}
\email[alexro@ciencias.unal.edu.co]{}




\affiliation{Departamento de F\'{\i}sica, Universidad Nacional de Colombia\\
Bogota, Colombia}


\begin{abstract}
A supersymmetric version of the $SU(3)_L \otimes U(1)_X$ gauge model is presented. The model is embedded in a $SU(6)_L \otimes U(1)_X$ gauge group and it has been built up completely chiral anomaly free and without any exotic charges in the fermionic content. The 
Minimal Supersymmetric Standard Model can be seen as an effective theory of this larger symmetry. We analyze how the spectrum of the new model can shift the upper bound on the CP-even Higgs boson and we find that it can be moved up to 143 GeV.
\end{abstract}
		
\maketitle

\section{Introduction}

The Standard Model (SM) has been established with great precision as the model that describe interactions of the gauge bosons with the 
fermions \cite{sm}. However, the Higgs sector, responsible for the symmetry breaking  has not been tested yet. But in any case the SM still 
has many unanswered questions and there are some experimental results on solar and atmospheric neutrinos which suggest no standard interpretation. The observation of neutrino oscillations can be explained by introducing mass terms for these particles, in whose case the mass eigenstates are different from the interaction eigenstates, it should be noticed that the oscillation phenomena imply that lepton family number is violated and such fact leads to consider the existence of physics beyond of the SM \cite{neutrinos}.

Among the various candidates to physics beyond the SM, supersymmetric theories play a special role. 
Supersymmetric models are the most developed and most widely accepted. Although there is not yet direct experimental evidence of 
Supersymmetry (SUSY), there are many theoretical arguments indicating that SUSY might be of relevance for physics beyond the SM. The main reason behind the phenomenological interest is that SUSY provides a solution for the naturalness problem which allows a fundamental scalar of the order of the electroweak scale if the mass of the superpartners are of the order of $\sim \cal O$(1 TeV).
The most popular version, of course, is the supersymmetric version of the SM, usually called MSSM, which is a computable theory and 
it is compatible with all precision electroweak tests \cite{precision}. 

Up to know, one impressive phenomenological virtue in favor 
of SUSY is obtained from the unification of couplings  in Grand Unified Theories, but on the other hand the lack of SUSY 
signals at LEP and the lower limit on $m_h$ pose problems for the MSSM \cite{search, hunter}. About this point, unlike in the SM, in the MSSM the coefficient  of the quartic potential of the Higgs sector is not an independent parameter, instead it is related with the ratio $\tan \beta$ and the $SU(2)_L \otimes U(1)_Y$ gauge couplings. As a consequence, the value of the CP-even scalar mass is not controlled by the SUSY breaking parameters but by the electroweak breaking sector. This yields the tree level bound $m_h^2 \leq m_Z^2 \cos^2 2 \beta \leq m_Z^2$, mass range already excluded by LEPII and could rule out the MSSM. But if the leading effects included in radiative corrections from the top-stop sector are taken into account, the upper bound can be pushed up to 135 GeV \cite{bound}. Now, the limit coming from LEP on the SM Higgs is $m_h \geq 114$ GeV \cite{lep2}, putting  restrictions on the parameter space of the MSSM, strong enough to consider new ideas beyond the SM that can shift this bound. The physical Higgs mass can be increased by enhancing the quartic coupling through an extended gauge sector and/or new superpotential Higgs couplings, in other words through D-terms or F-terms \cite{espinosa}. 

Among the possibilities with the nice properties of SUSY theories and the possibility to relax the bound on the lightest Higgs boson is
$SU(3)_C \otimes SU(3)_L \otimes U(1)_X$ as a local gauge theory.  It has been studied previously by many authors in non-SUSY and 
SUSY versions, who have explored different spectra of fermions and Higgs bosons \cite{31, pleitez, nos}.  There are many considerations 
about  this model but the most studied motivations of this large symmetry are the possibility to give mass to the neutrino sector \cite{pleitez}, 
anomaly  cancellations in a natural way in the 3-family version of the model, and an interpretation of the number of the fermionic 
families related with the anomaly cancellations \cite{su3}. 
Recently, reference \cite{romart} has presented a careful analysis of these kind of models without SUSY, taking into account 
the anomaly cancellation constraints. In fact, the version called model A in reference \cite{romart} was already supersymmetrized 
\cite{nos} and  it  was showed to be an anomaly free 
model, and a  family independent theory. Reference \cite{nos} has also shown that the SUSY models based on the symmetric $SU(3)_L \otimes U(1)_X$ model 
 as a $E_6$ subgroup can shift the upper bound on the lightest CP-even higgs boson up to 140 GeV. In the present work, we supersymmetrize the version 
called model B in reference \cite{romart}. The models A and B are different being the first one
an $E_6$ subgroup in contrast to the second one which is a $SU(6)_L\otimes U(1)_X$ subgroup.

The model presented here is a supersymmetric version of the gauge symmetry $SU(3)_L \otimes U(1)_X$ but it is different from the versions considered previously \cite{ma, nos}. The model considered  does not introduce Higgs triplets in the spectrum to break the symmetry, instead they are included 
in the scalar sector of the chiral superfields.  As we will show, the free parameters of the model is also reduced by using a basis where only one of 
the Vacuum Expectation Values (VEV) of the neutral singlets of the spectrum generates the breaking of the larger symmetry to the SM symmetry.  
Moreover, the fermionic content presented here does not have any exotic charges. Finally, our aim is to study how the spectrum of the new model based on the symmetry  $SU(3)_L \otimes U(1)_X$ can change the upper bound on the lightest CP-even Higgs boson.

This work is organized as follows. In section 2, we present the details of the non-SUSY version of the  $SU(3)_L \otimes U(1)_X$ model. In section 3 we discuss the SUSY version and the spontaneous symmetry breaking mechanism,  as well as some phenomenological implications of the model. Section 4 contains our conclusions.

\section{Non-SUSY version}

We want to describe the supersymmetric version of the $SU(3)_C\otimes SU(3)_L\otimes U(1)_X$ gauge symmetry which is embedded into 
$SU(6)_L\otimes U(1)_X$ \cite{romart}. But in order to be clear, first of all we present the non-SUSY version of the model. 
We assume that the left handed quarks (color triplets) and
left-handed leptons (color singlet) transform as the $\bar{3}$ and $3$
representations of $SU(3)_L$ respectively. In this  model the anomalies cancel individually in each family as it is done in the SM.

The model thus ends up with the following anomaly free multiplet structure:
\begin{eqnarray}
\begin{array}{||c|c|c|c||}\hline\hline
\chi_L=\left(\begin{array}{c}d\\u\\U \end{array}\right)_L & d^c_L & u^c_L&
U^c_L \\ \hline (3,\bar{3},{1\over 3}) & 
(\bar{3},1,{1\over 3}) & (\bar{3},1,-{2\over 3})
& (\bar{3},1,-{2\over 3}) \\ \hline\hline \end{array} 
\end{eqnarray}

\begin{eqnarray}
\begin{array}{||c|c|c||}\hline\hline
\psi_{1L}=\left(\begin{array}{c} \nu_e\\ e^-\\E^-_1\end{array}\right)_L & 
\psi_{2L}=\left(\begin{array}{c}
N_2^0\\E^-_2\\E^-_3\end{array}\right)_L &
\psi_{3L}=\left(\begin{array}{c} E^+_2\\ N_1^0\\ \nu^c_e\end{array}
\right)_L 
\\ \hline
(1,3,-{2\over 3}) & (1,3,-{2\over 3}) & (1,3,{1\over 3})
 \\ \hline\hline
\end{array}  
\end{eqnarray}

\begin{eqnarray}
\begin{array}{||c|c|c||}\hline\hline
 e^+_L & E_{1L}^+ & E_{3L}^+ \\ \hline
(1,1,1) & (1,1,1) & (1,1,1) \\ \hline \hline
\end{array} 
\end{eqnarray}
The numbers in parenthesis refer to the ($SU(3)_C, SU(3)_L, U(1)_X$) 
quantum numbers respectively. The symmetry is broken following the pattern 
\[SU(3)_C\otimes SU(3)_L\otimes U(1)_X\longrightarrow SU(3)_C\otimes SU(2)_L\otimes
U(1)_Y\longrightarrow SU(3)_C\otimes U(1)_Q\] 
\noindent
and give masses to the fermion fields in the model. 
With this in mind, it is necessary to introduce the following two Higgs triplets: $\phi_3(1,3,1/3)$ with a vacuum expectation value (VEV) aligned in the direction $\langle\phi_3\rangle=(0,0,V)^T$ and $\phi_4(1,3,1/3)$ with a VEV aligned as $\langle\phi_4\rangle=(0,v/\sqrt{2},0)^T$, with the hierarchy 
$V >v\sim 250$ GeV (the electroweak breaking scale). Note that $\phi_3$ and $\phi_4$ have the same quantum numbers but they get VEV's at 
different mass scales generating a large mass to the exotic $U$ quark and a realistic mass to the known up quark. 
One more Higgs scalar $\phi_2(1,3,-2/3)$ is needed in order to give a mass 
to the down quark field in the model. With the same scalar fields the neutral and charged fermions get masses through the see-saw mechanism \cite{seesaw}.
Diagonalizing the mass matrices there are a light Dirac neutrino and a heavy Dirac neutrino. Also there are a light charged lepton which corresponds to electron and three exotic heavy charged leptons.

The model has 9 gauge bosons. One of them
$B^\mu$ is associated with $U(1)_X$ symmetry,  and other 8 fields  are associated
with the $SU(3)_L$ symmetry. The expression for the electric charge generator in  $SU(3)_L \otimes U(1)_X$ is a linear combination of the three diagonal generators of the gauge group
\begin{equation}
Q=T_{3L}+\frac{1}{\sqrt{3}}T_{8L}+XI_3
\end{equation}
where $T_{iL}=\lambda_i/2$ with $\lambda_i$ the Gell-Mann matrices and $I_3$ the unit matrix.

After breaking the symmetry, we get mass terms for the charged and the neutral gauge bosons. By diagonalizing the matrix of the neutral gauge bosons we get the physical mass eigenstates which are
defined through the mixing angle $\theta_W$ given by
$\tan \theta_W=\sqrt{3}g_1/\sqrt{3g^2+g_1^2}$  with $g_1$ and $g$ the coupling constants of the $U(1)_X$ and $SU(3)_L$ respectively. 
Also we can identify the $Y^\mu$ hypercharge field associated
with the SM gauge boson as:
\begin{equation}
Y^\mu=\frac{\tan \theta_W}{\sqrt{3}}A_8^\mu+
(1-\tan \theta_W^2/3)^{1/2}B^\mu.
\end{equation}
In the SM the coupling constant $g'$ associated with the hypercharge $U(1)_Y$,
can be given by $\tan \theta_W=g'/g$ where $g$ is the coupling constant of $SU(2)_L$ which in turn can be taken equal to the $SU(3)_L$ coupling constant. Using the $\tan \theta_W$ given by the diagonalization of the neutral gauge boson matrix we obtain the matching condition
\begin{equation}\label{gut}
\frac{1}{g'^2}=\frac{1}{g_1^2}+\frac{1}{3g^2}.
\label{matching}
\end{equation}
We shall use this relation to write $g_1$ as a function of $g'$ in order to find the Higgs potential of the $SU(3)\otimes U(1)_X$ SUSY model at
low energies and compare it with the MSSM one. In particular, we will show that it reduces to the MSSM in this limit.

\section{SUSY version}

In the SUSY version the above content of fermions should be  written in terms of chiral superfields, and the gauge fields will be in vector supermultiplets as it is customary in SUSY theories. In this case we should introduce two more fermion fields and one more scalar. They will be arranged in the corresponding superfields, in such way that they cancel chiral anomalies automatically, and additionally the Higgs scalar fields break the symmetry and generate the fermion and gauge boson masses. Therefore the contents of superfields are
\begin{eqnarray}
\begin{array}{||c|c|c|c||}\hline\hline
\hat \chi_L=\left(\begin{array}{c}\hat d\\\hat u\\\hat U \end{array}\right)_L & \hat d^c_L & \hat u^c_L&
\hat U^c_L \\ \hline (3,\bar{3},{1\over 3}) & 
(\bar{3},1,{1\over 3}) & (\bar{3},1,-{2\over 3})
& (\bar{3},1,-{2\over 3}) \\ \hline\hline \end{array} 
\label{contf1}
\end{eqnarray}

\begin{eqnarray}
\begin{array}{||c|c|c||}\hline\hline
\hat \psi_{1L}=\left(\begin{array}{c} \hat \nu_e\\ \hat e^-\\\hat E^-_1\end{array}\right)_L & 
\hat \psi_{2L}=\left(\begin{array}{c} \hat N_2^0\\\hat E^-_2\\\hat E^-_3\end{array}\right)_L &
\hat \psi_{3L}=\left(\begin{array}{c} \hat E^+_2\\ \hat N_1^0\\ \hat \nu^c_e\end{array}\right)_L  \\ \hline
(1,3,-{2\over 3}) & (1,3,-{2\over 3}) & (1,3,{1\over 3}) 
 \\ \hline\hline
\end{array} 
\label{contf2} 
\end{eqnarray}

\begin{eqnarray}
\begin{array}{||c|c|c|c|c||}\hline\hline
\hat \psi_{4L}=\left(\begin{array}{c} \hat E^+_4\\ \hat N_4^0\\ \hat N_5^0 \end{array}\right)_L &
\hat \psi_{5L}=\left(\begin{array}{c} \hat E^-_4\\ \hat N_7^0\\\hat N^0_6 \end{array}\right)_L &  
\hat e^+_L & \hat E_{1L}^+ & \hat E_{3L}^+ \\ \hline
(1,3,{1 \over 3})  &  (1, \bar 3, -\frac 1 3)& (1,1,1) & (1,1,1) & (1,1,1)\\ \hline \hline
\end{array} 
\label{contf3}
\end{eqnarray}
The eqs. (\ref{contf1}) and (\ref{contf2}) plus singlets of equation (\ref{contf3}) have the same fermionic content as the non-SUSY version which is anomaly free. In order to give masses in the non-SUSY model it was necessary to introduce a scalar Higgs $\phi_4$. To supersymmetrize the model we have introduced a chiral superfield with the same quantum numbers $\hat \psi_{4L}$. But it introduces more fermions and in order to cancel the triangle anomalies a new chiral superfield is necessary $\hat \psi_{5L}$, in such way that 
$\hat \psi_{4L}$ and $\hat \psi_{5L}$ are in a vector representation of the $SU(3)_L\otimes U(1)_X$ gauge symmetry and then the model is completely chiral anomaly free.

The scalar sector of the supermultiplets has the appropriate quantum numbers in order to be useful as a Higgs sector of the theory and is not necessary to introduce new scalar fields to get the spontaneous symmetry breaking to the SM symmetry. The superpotential can be built up by taking into account the non-SUSY Yukawa Lagrangian form and  $\epsilon_{abc} \psi^a_i \psi^b_j \psi^c_k$ which is a singlet; we have
\begin{eqnarray}
W&=&h_1 \epsilon_{abc} \hat \psi_1^a \hat \psi_3^b \hat \psi_4^c +h_2 \epsilon_{abc} \hat \psi_3^a \hat \psi_2^b \hat \psi_4^c + 
         \hat \psi_1 (h_3 \hat \psi_5 \hat e^+ +h_4 \hat \psi_5 \hat E_1^+ + h_5 \hat \psi_5 \hat E_3^+) \nonumber \\
&+& \hat \psi_2 (h_6 \hat \psi_5 \hat e^+ +h_7 \hat \psi_5 \hat E_1^+ + h_8\hat \psi_5 \hat E_3^+)+ h_u \hat \chi_L \hat \psi_4 \hat u + 
h_U \hat \chi_L \hat \psi_4 \hat U + h'_u \hat \chi_L \hat \psi_3 \hat u \nonumber \\
&+& h'_U \hat \chi_L \hat \psi_3 \hat U +  h_d \hat \chi_L \hat \psi_2 \hat d + 
h'_d \hat \chi_L \hat \psi_1 \hat d + \mu_1 \hat \psi_4 \hat \psi_5 + \mu_2 \hat \psi_3 \hat \psi_5 .      
\end{eqnarray}
Once we have the superpotential $W$, the theory is defined and we can get the Yukawa interactions and the scalar potential. Here, we should mention that the lepton number is not conserved, in order to generate neutrino masses similar to $R_p$ violating models \cite{losada}.

Now, we will concentrate our attention on  the scalar Higgs potential, which is given by
\begin{equation}
V=\left | \frac {\partial W}{\partial A_i} \right |^2 + \frac 1 2 D^a D^a + \frac 1 2 D' D'
\end{equation}
where
\begin{eqnarray}
D^a &=& g A_i^{\dagger} \frac{\lambda_{ij}^a}{2} A_j \; \nonumber ,\\
D' &=& g_1 A_i^{\dagger} X(A_i) A_i  
\end{eqnarray}
and $A_i$ are the scalar components of the chiral supermultiplets. For our purpose, it is enough to consider only
\begin{equation}
W'=h_1 \epsilon_{abc} \hat \psi_1^a \hat \psi_3^b \hat \psi_4^c +h_2 \epsilon_{abc} \hat \psi_3^a \hat \psi_2^b \hat \psi_4^c + \mu_1 \hat \psi_4 \hat \psi_5 + \mu_2 \hat \psi_3 \hat \psi_5.       
\end{equation}
As we already mentioned, the scalar Higgs bosons responsible of the symmetry breaking are the scalar sector of the fermionic chiral supermultiplets, thus we have $\phi_i=\tilde \psi_{iL}$. Now we focus on the scalar potential obtained from the superpotential $W'$ plus the soft breaking terms which are 
\begin{equation}
V_{soft}=\sum_{i=1}^5 m_i^2 \phi_i^\dagger \phi_i +\mu'_1(\phi_4^\dagger \phi_5 + h.c)+\mu'_2 (\phi_3^\dagger \phi_5 + h.c).
\end{equation}
The D-terms and F-terms are
\begin{eqnarray}
V&&=\frac{g^2}{8}\{ \frac 43 \sum_{a=1}^5 (\phi_a^\dagger \phi_a)^2 +4 \sum_{a\neq b}^5 (\phi_a^\dagger \phi_b)(\phi_b^\dagger \phi_a)-\frac 43 \sum_{a \neq b}^5 (\phi_a^\dagger \phi_a)(\phi_b^\dagger \phi_b)\} \nonumber \\
&+& \frac{g_1^2}{18} \{ -2(\phi_1^\dagger \phi_1+\phi_2^\dagger \phi_2)+(\phi_3^\dagger \phi_3+\phi_4^\dagger \phi_4)-\phi_5^\dagger \phi_5\}^2 \nonumber \\
&+& h_1^2 \{(\phi_3^\dagger \phi_3)(\phi_4^\dagger \phi_4)+(\phi_1^\dagger \phi_1)(\phi_3^\dagger \phi_3)-(\phi_3^\dagger \phi_4)(\phi_4^\dagger \phi_3)-(\phi_1^\dagger \phi_3)(\phi_3^\dagger \phi_1)  \} \nonumber \\
&+& h_2^2 \{(\phi_3^\dagger \phi_3)(\phi_4^\dagger \phi_4)+(\phi_2^\dagger \phi_2)(\phi_3^\dagger \phi_3)-(\phi_3^\dagger \phi_4)(\phi_4^\dagger \phi_3)-(\phi_2^\dagger \phi_3)(\phi_3^\dagger \phi_2)  \} \nonumber \\
&+& h_1^2 \{ (\phi_1^\dagger \phi_1)(\phi_4^\dagger \phi_4)- (\phi_1^\dagger \phi_4) (\phi_4^\dagger \phi_1)\}+h_2^2\{ (\phi_2^\dagger \phi_2) (\phi_4^\dagger \phi_4)- (\phi_2^\dagger \phi_4) (\phi_4^\dagger \phi_2)\} \nonumber \\
&+& 2 h_1 h_2 \{ (\phi_1^\dagger \phi_2) (\phi_4^\dagger \phi_4)- (\phi_1^\dagger \phi_4) (\phi_2^\dagger \phi_4)+ (\phi_1^\dagger \phi_2) (\phi_3^\dagger \phi_3)- (\phi_1^\dagger \phi_3) (\phi_2^\dagger \phi_3) \} \nonumber \\
&+& (\mu_1^2 +\mu_2^2) (\phi_5^\dagger \phi_5)+\mu_1^2  (\phi_4^\dagger \phi_4)+ \mu_2^2  (\phi_3^\dagger \phi_3)+ \mu_1 \mu_2  (\phi_4^\dagger \phi_3+ \phi_3^\dagger \phi_4).
\label{potencial}
\end{eqnarray}

Now, we are ready to break down the symmetry  $SU(3)_L \otimes U(1)_X$ to the SM symmetry  $SU(2)_L \otimes U(1)_Y$. Writting the Higgs triplets of $SU(3)_L$ as  branching rules of $SU(2)_L$, we have
\begin{eqnarray}
\begin{array}{ccc}
\phi_1=\left(\begin{array}{c}\tilde  l\\\tilde E_1^- \end{array}\right), & 
\phi_2=\left(\begin{array}{c} H_1 \\\tilde E_3^- \end{array}\right),  & 
\phi_3=\left(\begin{array}{c} H_2 \\ \tilde N_1^0 \end{array}\right), \\ 
\phi_4=\left(\begin{array}{c} H_4 \\\tilde N_5^0 \end{array}\right), & 
\phi_5=\left(\begin{array}{c} H_5 \\\tilde N_6^0 \end{array}\right), \\  
\end{array} 
\end{eqnarray}
where $\tilde l$ and $H_i$ are doublets of $SU(2)_L$ and $\tilde E^-_i$, $\tilde N_i^0$ are singlets. Thus the VEV's of
$\langle \tilde N_1^0 \rangle = u_1$, $\langle \tilde N_6^0 \rangle =u_2$ and $\langle \tilde N_5^0 \rangle =u$
make the job. But it is possible to choose only one of them different from zero, $u_1=u_2=0$, $u\neq 0$ \cite{georgi}, and the would-be Goldstone 
bosons of the symmetry breaking $SU(3)_L \otimes U(1)_X \big/ SU(2)_L\otimes U(1)_Y$  become degrees of freedom of the field $\phi_4$. 
The terms involving $H_1$, $H_2$ doublets generate the fermion masses in the 
non-SUSY model after the electroweak symmetry breaking. Since we want to generate the MSSM at low energy as an effective theory we suppose that 
$\phi_1$, $\phi_5$ and $\tilde N_1^0$ are decoupled at low energies too.  Therefore we only deal with a Higgs potential which involves $H_1$, $H_2$ and $\tilde N_5^0$. It is given by
\begin{eqnarray}
V' &&=\frac{g^2}{8}\{ \frac 43 ((H_1^\dagger H_1)^2 + (H_2^\dagger H_2)^2 +(\tilde N_5^2)^2) +4(H_1^\dagger H_2)(H_2^\dagger H_1)\nonumber \\
&-&\frac 43((H_1^\dagger H_1)(H_2^\dagger H_2)+(H_1^\dagger H_1)\tilde N_5^2 +(H_2^\dagger H_2)\tilde N_5^2) \} \nonumber \\
&+& \frac{g_1^2}{18} \{ \tilde N_5^2 +(H_2^\dagger H_2)-2(H_1^\dagger H_1)\}^2 + h_1^2 (H_2^\dagger H_2) \tilde N_5^2 \nonumber \\
&+& h_2^2 \{(H_2^\dagger H_2)\tilde N_5^2 +(H_1^\dagger H_1)(H_2^\dagger H_2)-(H_1^\dagger H_2)(H_2^\dagger H_1)+(H_1^\dagger H_1)\tilde N_5^2 \} \nonumber \\
&+& \mu_1^2 \tilde N_5^2 + \mu_2^2 (H_2^\dagger H_2)+m_1^2(H_1^\dagger H_1)+m_2^2(H_2^\dagger H_2)+m_5^2 \tilde N_5^2.
\end{eqnarray}

As it has been already emphasized \cite{ma}, in the MSSM the quartic scalar couplings of the Higgs potential are completely determined in 
terms of the two gauge couplings, but it is not the case if the symmetry $SU(2)_L \otimes U(1)_Y$  is a remnant of a larger symmetry which 
is broken at a higher mass scale together with the SUSY. The scalar particle content needed to produce the spontaneous symmetry breaking 
determines the structure of the Higgs potential. In this way, the reduced Higgs potential would be a 2HDM-like potential, but its quartic 
couplings would not be those of the MSSM, instead they will be related to the gauge couplings of the larger theory and to the couplings 
appearing in its superpotential. Analyses of supersymmetric theories  in this context  have been given in the literature \cite{ma,nos,others}. 
In particular, it has been studied widely for different versions of the left-right model, SUSY version of the $SU(3)_L \otimes U(1)_X$ model where 
exotic charged particles of electric charges $(-4/3, 5/3)$ appear \cite{georgi} and the version where $SU(3)_L \otimes U(1)_X$ is an $E_6$ 
subgroup \cite{nos}. 

Following this idea with the reduced Higgs potentials presented in the previous paragraph, we can obtain the effective quartic scalar couplings $\lambda_i$ of the most general 2HDM potential.  Since there are cubic interactions in $V'$ involving $H_{1,2}$ and $\tilde N_5^0$, it generates two types of Feynman diagrams which contribute to the quartic couplings \cite{nos}. The Feynman rules from the potential for these couplings are
\begin{eqnarray}
H_1-H_1-\tilde N_5^0 &\to&  i(2h_2^2 -\frac{1}{9}(3g^2+4g_1^2))u , \nonumber \\
H_2-H_2-\tilde N_5^0 &\to&  i(2(h_1^2+h_2^2) -\frac{1}{9}(3g^2-2g_1^2))u
\end{eqnarray}
and using them we obtain the effective couplings; thus they are given by
\begin{eqnarray}
\frac{\lambda_1}{2} &=& \frac{2 g_1^2}{9}+ \frac{g^2}{6}-\frac 3 8 
\left((\frac{4 g_1^2}{9}+\frac{g^2}{3})^2- 4 h_2^2(\frac{4 g_1^2}{9}+\frac{g^2}{3}) + 4h_2^4 \right) G^{-1} \;\; , \nonumber \\
\frac {\lambda_2}{2} &=& \frac{g_1^2}{18}+ \frac{g^2}{6}-\frac 3 8 
\left((\frac{2 g_1^2}{9}-\frac{g^2}{3})^2+4 (h_1^2+h_2^2)(\frac{2 g_1^2}{9}-\frac{g^2}{3}) + 4 (h_1^2 + h_2^2)^2 \right) G^{-1} \; \; ,\nonumber \\
\lambda_3 &=& -\frac{2 g_1^2}{9}- \frac{g^2}{6}+\frac 3 4 
\left((\frac{4 g_1^2}{9}+\frac{g^2}{3})(\frac{2 g_1^2}{9}- \frac{g^2}{3})- 4 h_2^2 (h_1^2+h_2^2) + 2 h_2^2(\frac{g^2}{3}-\frac{2g_1^2}{9}) \right. \nonumber \\
&+& \left. 2(h_1^2+h_2^2)(\frac{g^2}{3}+
\frac{4 g_1^2}{9}) \right) G^{-1} \; \; ,\nonumber \\
\lambda_4&=&\frac{g^2}{2}-h_2^2 \; \; \; , \; \; \nonumber \\
\lambda_5&=&0 
\end{eqnarray}
where $G=(g_1^2/3+g^2)$. Additionally, we realize that in the limit of $h_1=0$ and $h_2=h_e$ we recover the $SU(3)_L\otimes U(1)_X$ model previously worked out \cite{nos}.

We want to remark that 
this SUSY model has the MSSM as an effective theory when the new physics is not longer there, that means
$h_i=0, i=1, 2$, and the coupling constants are running down to the electroweak scale.  At this point we use the 
approach where the $SU(2)_L$ coupling behaves like $g$. 
In the limit $h_1=h_2=0$, we obtain
\begin{equation}
\lambda_{1,2} = \frac {g^2(4g_1^2+3g^2)}{4(g_1^2+3g^2)} \; \; , \; \;
\lambda_3 = - \frac {g^2(4g_1^2+3g^2)}{4(g_1^2+3g^2)} \, \, , \,\, 
\lambda_4=\frac{g^2}{2} \nonumber
\end{equation}
and, if we assume the matching condition from equation (\ref{gut}), we reduce the effective couplings to those appearing in the MSSM, as expected,
\begin{equation}
\lambda_1=\lambda_2=\frac 14 (g^2+g'^2) \; \; ,\;\;
\lambda_3= - \frac 14 (g^2+g'^2) \, \, , \, \, \lambda_4 =\frac{g^2}{2}. \nonumber
\end{equation}

But when new physics is present, the conditions coming from the analysis of the Higgs potential on the new parameters $h_i$ are quite different. The requirement that the potential be bounded from below, implies that $\lambda_1 \geq 0$, $\lambda_2 \geq 0$ and either $\lambda_3 +\lambda_4 \geq 0$ or if $\lambda_3+\lambda_4<0$ then $\vert \lambda_3 +\lambda_4 \vert \leq \sqrt{\lambda_1 \lambda_2}$ ,  this is not a simple region, and therefore in order to get a bound on the lightest Higgs boson, we evaluate numerically point by point the conditions, and the results are shown in figure 1. Furthermore, we have added the term coming from radiative corrections due to the top quark and its two supersymmetric partners \cite{bound, espinosa, ma}, taking $m_t=174.3$ GeV \cite{pdg} and $\tilde m=1$ TeV. Figure 1 is a 3D-plot for three different values of $\cos^2 \beta$ where $\tan \beta \equiv \langle H_2 \rangle_0/ \langle H_1 \rangle_0$, and the lowest bound of MSSM can be moved up to 143 GeV. The range of values for $h_1^2$ and $h_2^2$ are presented in such a way that they satisfy the inequalities coming from the requirement that the potential be bounded from below, and we note that for $h_1^2=0$ the largest value for $h_2^2$ is 0.21.  
 
On the other hand, the upper bound of the Higgs boson mass has been calculated driving the VEV's of the SM singlets in such way that only one VEV has been included, but we can explore the behavior of this bound considering more singlets in the model which have couplings in the same form that the first one $\tilde N_5^0$. Doing that, we note that the upper bound of the lightest Higgs boson is not affected pretty much, but the constraints on the parameters $h_i$, are. In figure 2, we show the plane $h_1^2-h_2^2$ considering one, three and ten singlets in the model. The figure implies that the new physics related to the parameters $h_i$ is more constrained when the number of singlets involved increases.

\section{CONCLUSIONS}
We have presented a new supersymmetric version of the symmetry $SU(3)_L \otimes U(1)_X$ which is a $SU(6)_L \otimes U(1)_X$ subgroup, where the Higgs bosons are in the scalar sector of the superfields, 
and it is triangle anomaly free. We have also 
shown that using the limit when the parameter $h_i=0$ and  the matching condition (equation (6)), we obtain the SUSY constrains 
for the Higgs potential as in the MSSM. Therefore, if we analyze the upper bound for the mass of the lightest CP-even Higgs boson in this 
limit, we find the 
same bound of around 128 GeV for the MSSM. However, since in general $h_i \not = 0$ $i=1, 2$, such upper bound can be moved up to around 143 GeV. 
This fact can be an interesting alternative to take into account in the search for the lightest CP-even SUSY Higgs boson mass. This model is different from the one considered in  \cite{nos} and only 
can be recovered  when $h_1=0$ and $h_2=h_e$. By considering the experimental value from LEP for the lightest Higgs mass $m_h\geq 114$ GeV it is possible to get bounds for the parameters of the model $ h_i^2$, they are $0\leq h_1^2 \leq 0.21$, $0 \leq h_2^2 \leq 0.21$. But if we include more singlets of the SM symmetry like $\tilde N_5^0$, the upper bound on the Higgs boson mass does not move considerably, only up to 145 GeV, but the allowed values in the plane $h_1^2-h_2^2$ does. These allowed values are reduced when the number of singlets added are increased.

We acknowledge to R. Diaz for the careful reading of the manuscript. This work was supported by COLCIENCIAS and DIB.

\newpage

\begin{figure}[htbp]
 {\hbox{
    \includegraphics[width=10cm, angle=-90]{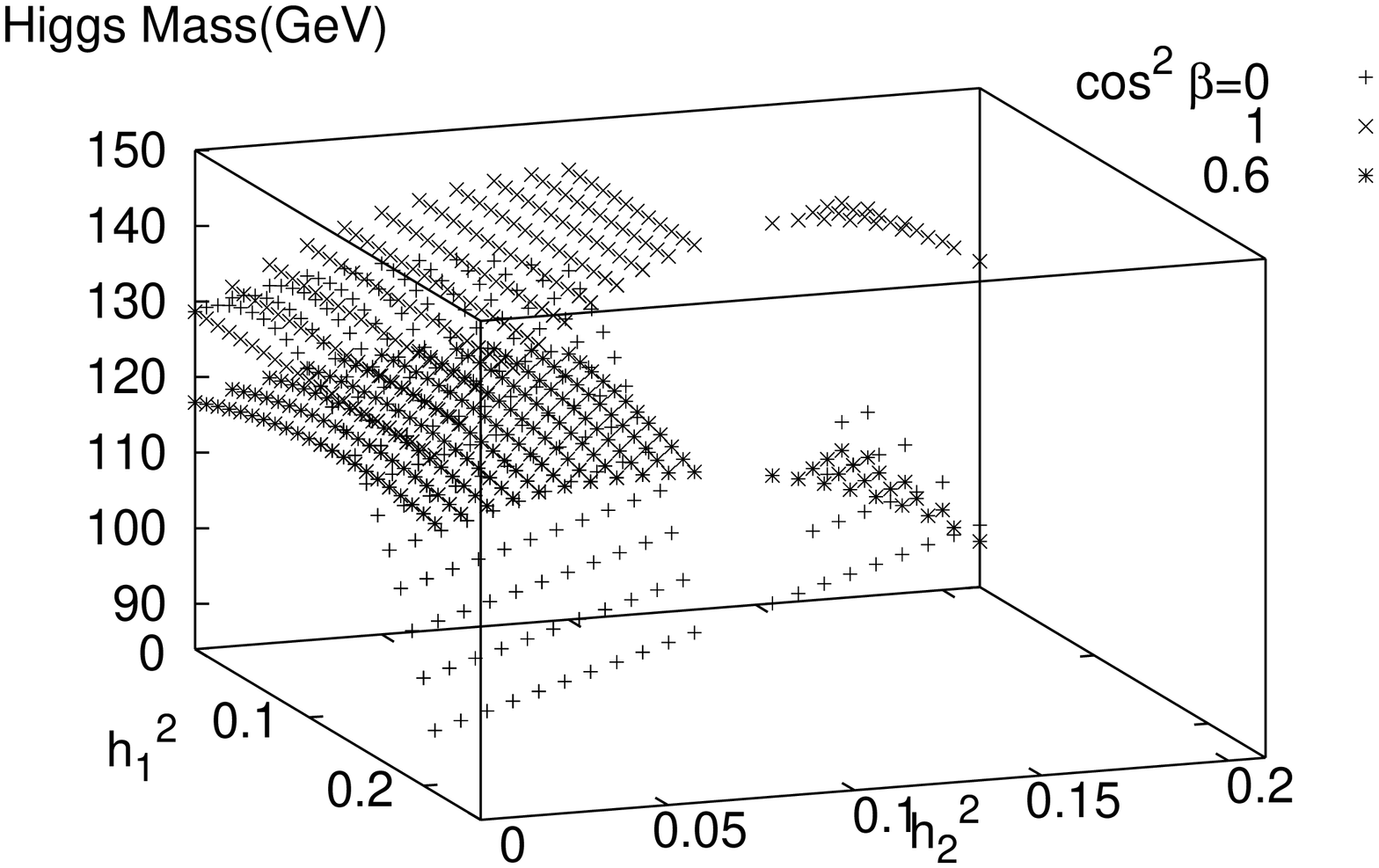} }}
 \caption{\it The upper bound on the lightest CP-even Higgs boson as a function of the parameters $h_i^2$ from the superpotential. 
The  different kind of points  correspond to $\cos^2 \beta =\{0\;,\;0.6\;,\;1\}$. }   
\end{figure}

\begin{figure}[htbp]
 {\hbox{
    \includegraphics[width=10cm, angle=-90]{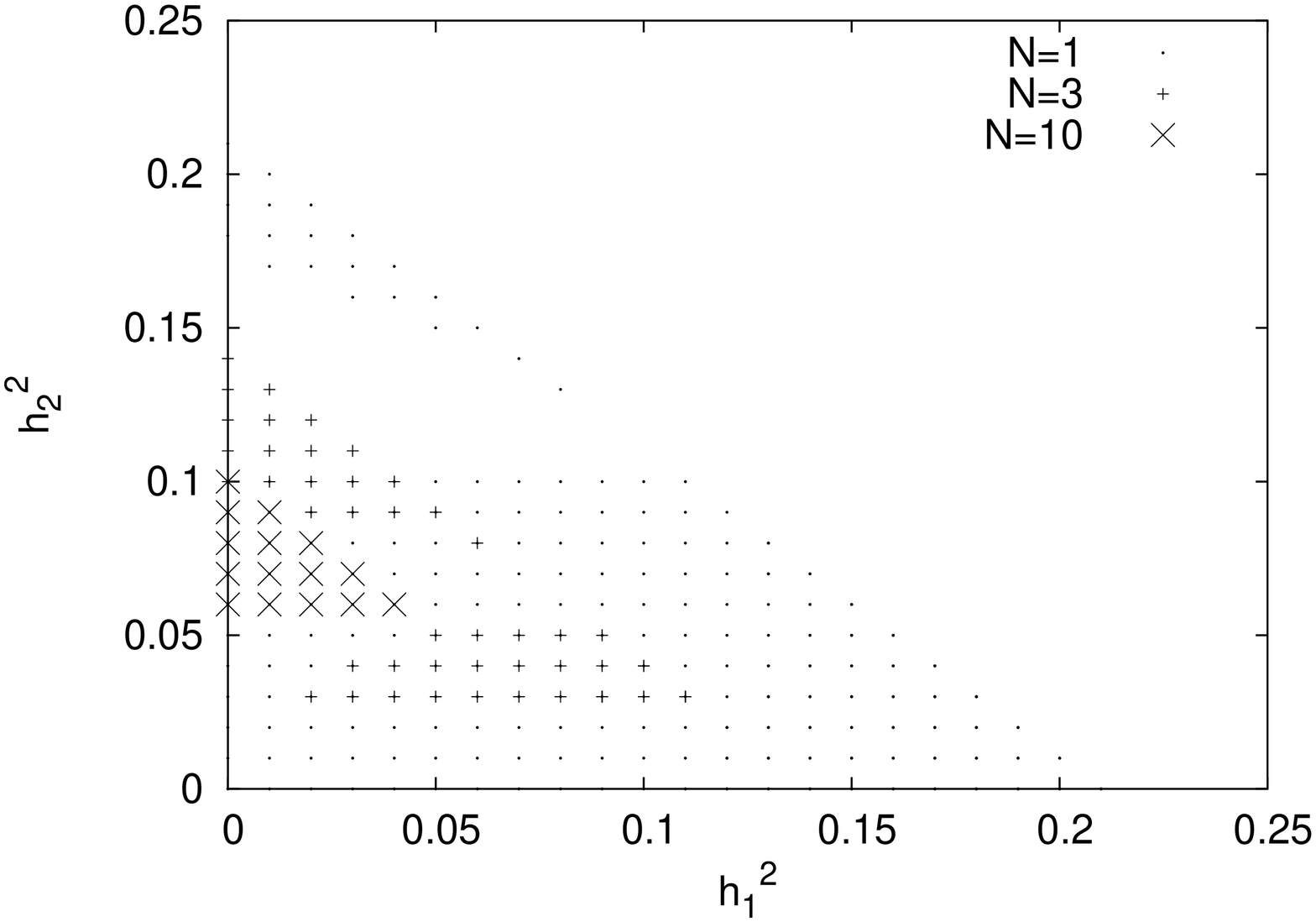} }}
 \caption{\it The allowed region for the plane $h_1^2-h_2^2$ with additional SM singlets, for $N=1$, $N=3$ and $N=10$.}   
\end{figure}

\begin{thebibliography}{000}

\bibitem{sm} S. Weinberg, Phys. Rev. Lett. 19 (1967) 1264; S. L. Glashow, Nucl. Phys. 20 (1961) 579; A. Salam, in: N. Svartholm (ed.), Elementary Particle Theory, 1968, p. 367.

\bibitem{neutrinos}Beshtoev Kh. M., arXiv:hep/0204324 and references therein.

\bibitem{precision} S. Heinemeyer, G. Weiglein, JHEP 0210 (2002) 072; arXiv:hep-ph/0307177, arXiv:hep-ph/0301062; P. Gambino, arXiv:hep-ph/0202001; U. Baur, et. al., eConf C010630 (2001)P1WG1, report of the electroweak precision measurements working group at Snowmass 2001.
 
\bibitem{hunter} J. F. Gunion, H. Haber, G. Kane, S. Dawson, The Higgs hunter's guide, Addison-Wesley, Redwood city, 1990.

\bibitem{search} J. G. Branson, et. al., hep-ph/011021; D. Zeppenfeld, et. al., Phys. Rev. D 62 (2000) 0130009; 
A. Djouadi, W. Kilian, M. Muhlleitner and P.M. Zerwas, Eur. Phys. J. C 10 (1999) 45.

\bibitem{bound} M. Carena, S. Heinemeyer, C. E. Wagner, G. Weiglein, Eur. Phys. J. C 26 (2003)601; P.N. Pandita, Mod. Phys. Lett. A. 10 (1995) 1533; J. A Casas, J. R. Espinosa, M. Quiros, A. Riotto, Nucl. Phys. B 436 (1995) 3; G. L. Kane, C. Kolda, J. Wells, Phys. Rev. Lett. 70 (1993) 2686; H. Haber and R. Hemplfling, Phys. Rev. Lett. 66 (1991) 1815; J. Ellis, G. Ridolfi and F. Zwirner, 
Phys. Lett. B 257 (1991) 83. R. Barbieri, M. Frigeni and F. Caravaglios, Phys. Lett. B 258 (1991) 167.

\bibitem{lep2} LEPEWWG, hep-ex/0112021, http://lepewwg.web.cern.ch/LEPEWWG/ 

\bibitem{espinosa} J. R. Espinosa, M. Quiros, Phys. Lett. B 302 (1993) 51; A Brignole, J. A. Casas, J. R. Espinosa, I. Navarro, Nucl. Phys. B 666 (2003) 105; P. Batra, A. Delgado, D. Kaplan, T. M. P. Tait, arXiv:hep-ph/0309149.

\bibitem{31} C. H. Albright, C. Jarlskog and M. Tjia, Nucl. Phys. B86 (1974) 535 and references there in. F. Pisano and 
V. Pleitez, Phys. Rev. D 46 (1992) 410.


\bibitem{pleitez}F.Pisano and V. Pleitez, Phys. Rev. D46, 410 (1992); P. H. Frampton, Phys. Rev. Lett. 69, 2889 (1992); R. Foot, O. F. Hernandez, F. Pisano and V. Pleitez, Phys. Rev. D47, 4158 (1993).


\bibitem{nos}Rodolfo Diaz, J. Mira, R. Martinez and J.-Alexis Rodriguez, Phys. Lett. B 552 (2003) 287.

\bibitem{su3} M. Singer, J. F. W. Valle and Schechter, Phys. Rev. D 22 (1980)738; R. Foot, H. N. Long and T.A. Tran, Phys. Rev. D 50, R34 (1994); 
V. Pleitez, Phys. Rev. D 53 (1996) 514.

\bibitem{romart} R. Martinez, W. Ponce and L. A. Sanchez, Phys. Rev. D 65 (2002) 055013; Phys. Rev D 64 (2001) 075013; W. Ponce, J. Florez and L. A. Sanchez, Int. J. Mod. Phys. A 17 (2002) 643.



\bibitem{ma} T. V. Duong and E. Ma, Phys. Lett. B 316 (1993) 307; E. Ma and D. Ng, Phys. Rev. D 49 (1994) 6164.

\bibitem{seesaw}T. Yanagida, in proc workshop on unified theory and baryon number in the universe, eds. O. Sawada and A. Sugamoto, (KEK 79).M. Gell-Mann, P. Ramond and R. Slansky, in supergravity eds. P. van Nieuwenhuizen and D. Freedman (North- Holland 1979); R. Mohapatra and G. Senjanovic, Phys. Rev. Lett. 44 (1980) 912.

\bibitem{losada} S. Davidson, M. Losada and N. Rius, Nucl. Phys. B 587 (2000) 118; A. Pilaftsis, M. Nowakowski, Nucl. Phys. B 461(1996)19; A.S. Joshipura, M. Nowakowski, Phys. Rev. D 51 (1995) 5271; E. Nardi, Phys. Rev. D 55 (1997) 5772.

\bibitem{georgi} H. Georgi and D. V. Nanopoulos, Phys. Lett. B 82 (1979), 1.


\bibitem{others} K. S. Babu, X. G. He and E. Ma, Phys. Rev. D 36 (1987) 878; E. Ma, Phys. Rev. D 36 (1987) 274.

\bibitem{pdg} K. Hagiwara, et. al.[Particle Data Group Collaboration], Phys. Rev. D 66 (2002) 0100001.

\end{thebibliography}
\end{document}